\documentclass{iopart}
\usepackage[dvips]{graphicx}

\begin{document}

\title{Quantum fluctuations and correlations of spatial scalar or multimode vector solitons in Kerr media}

\author{Eric Lantz\dag, Thibaut Sylvestre\dag, Herv\'{e} Maillotte\dag, Nicolas Treps\ddag, and Claude Fabre\ddag}

\address{\dag\ Laboratoire d'Optique P.M. Duffieux, Universit\'{e} de Franche-Comt\'{e}, U.M.R. CNRS 6603, 25030 Besan\c{c}on cedex, France}

\address{\ddag\ Laboratoire Kastler Brossel,
Universit\'e Pierre et Marie Curie, case 74, 75252 Paris cedex 05,
France.}


\begin{abstract} We apply the Green's function method to determine the
global degree of squeezing and the transverse spatial distribution
of quantum fluctuations of solitons in Kerr media. We show that
both scalar bright solitons and multimode vector solitons
experience strong squeezing on the optimal quadrature. For vector
solitons, this squeezing is shown to result from an almost perfect
anti-correlation between the fluctuations on the two
incoherently-coupled circular polarisations.
\end{abstract}

\maketitle

\section{Introduction}

\indent

 The analysis of spatial aspects of quantum fluctuations is now
a well established field of quantum optics (see for example
\cite{kolobov} for a review). More specifically, some recent
studies \cite{Nagasako,Mecozzi,Treps} have considered quantum
effects in spatial solitons. Ref. \cite{Nagasako} shows that
vacuum-induced jitter remains negligible versus the width of a
soliton. In Ref. \cite{Mecozzi}, production of subpoissonian light
is demonstrated by placing either a stop on the centre of the beam
in the near field or a low pass filter in the far-field. These two
theoretical papers use linearisation of quantum fluctuations. Ref.
\cite{Treps} gives a general method to numerically determine the
transverse distribution of linearised quantum fluctuations by
using Green functions. Results show an important global squeezing
for $\chi^2$  as well as for $\chi^3$  solitons. In Kerr media
however, this squeezing does not occur on the amplitude
quadrature, implying homodyne detection. Ref.\cite{Treps}
demonstrates also local squeezing and anti-correlations between
different parts of the beam. The Green's function method has also
been applied to the assessment of spatial aspects of quantum
fluctuations of spontaneous down-conversion \cite{Lantz}. In this
paper, we apply the Green's function method to spatial scalar
solitons and to multimode vector solitons in Kerr media
\cite{Kockaert,Cambournac}. The results for the scalar solitons
appear somewhat different of that given in ref. \cite{Treps},
because of a flaw in the numerics used in  this reference. We show
in particular that, in the corrected results, the squeezing
monotonically increases with the propagation length, like for a
plane wave. We show also that results of ref. \cite{Mecozzi} are
retrieved with the Green's function method.

We study also the quantum properties of multimode vector solitons
that result from incoherent coupling of two counter-rotating
circularly polarised fields mutually trapped through cross-phase
modulation, the self-focusing nature of a component being exactly
compensated by the diffracting nature of the other. We find the
quantum limits for the stability of these multimode vector
solitons and show that the global squeezing (measured on the total
intensity of the soliton) appears to be similar to that of scalar
solitons. In addition, we demonstrate that there exists a strong
anti-correlation between the circular components.

The paper is organised as follows. In Sec. 2, we recall the
principle of the Green's function method. Sec. 3 gives the
results for scalar Kerr solitons and Sec. 4  is devoted to vector
solitons.

\section{Green's function method}

Let us briefly recall the main features of the Green's function
method \cite{Treps} that allows us to determine the spatial
distribution of quantum fluctuations in the regime of free
propagation in a non-linear medium. We will present it in the
simplest case of a scalar field and in the configuration of a
single transverse coordinate x. The classical propagation equation
for the complex electric field envelope $U(x,z)$ is in this case :

\begin{equation}
\frac{\partial U}{\partial z}=\frac{i}{2k} \frac{\partial^2
U}{\partial x^2}+i\gamma|U|^2U\label{mean}
\end{equation}
where $\gamma$ is the usual Kerr coefficient, z the main
propagation direction, $k$ the longitudinal wavevector, and x the
transverse position in the transverse plane. This equation has a
solution that we will call $\bar{U}(x,z)$.

Let us write the quantum positive frequency operator describing
the same field at the quantum level, $\hat{U}(x,z)$, as :

\begin{equation}
\hat{U}(x,z) =  \bar{U}(x,z)+\delta\hat{U}(x,z) \label{quant}
\end{equation}
If the fluctuations remain small compared to the mean fields,
which is true as long as the field contains a macroscopic number
of photons, then the quantum mean field coincides with the field
given by classical nonlinear optics, $\bar{U}(x,z)$, and the
fluctuations $\delta\hat{U}$ obey a simple propagation equation,
obtained by linearizing the classical equation of propagation
\ref{quant} around the mean field. This equation writes :

\begin{equation}
\frac{\partial}{\partial z} \delta\hat{U}=\frac{i}{2k}
\frac{\partial^2}{\partial
x^2}\delta\hat{U}+i\gamma\left(2|\bar{U}|^2\delta\hat{U}+\bar{U}^2\delta\hat{U}^+\right)
\label{linear}
\end{equation}
As this equation is linear, one can use Green's function
techniques to solve it. Its solution at a given $z_{out}$ is a
linear combination of the input fluctuations operators in plane
$z_{in}$:

\begin{equation}
\delta\hat{U}(x,z_{out})=\int dx'
G(x,x')\delta\hat{U}(x',z_{in})+\int dx'
H(x,x')\delta\hat{U}^+(x',z_{in})
 \label{Green}
\end{equation}
where $G(x,x')$ and $H(x,x')$ are the Green's functions associated
with the linear propagation equation \ref{linear}. If the input
field fluctuations are those of a vacuum field, or of a coherent
field, they are such that :

\begin{equation}
<\delta\hat{U}(x,z_{in})\delta\hat{U}^+(x',z_{in})>=C\delta(x-x')
 \label{correl}
\end{equation}
where $C$ is a constant. Equations \ref{Green} and \ref{correl}
allow us to calculate quantum noise variance or correlation at the
output of our system on any quadrature component when one knows
the two Green's function $G(x,x')$ and $H(x,x')$ of the problem.
These two functions can readily be numerically calculated as
solutions of the propagation equation \ref{linear} when the
initial condition is a delta function localised at a given
transverse point. This method can be obviously extended to any
kind of non-linear propagation equation.

We will focus our attention here on two kinds of simple spatially
resolved  measurements : the first one is the direct
photodetection on a pixellised detector, which gives access to the
local intensity fluctuations ; the second one is the spatially
resolved balanced homodyne detection, using a local oscillator
having a plane wavefront, coincident with the detector plane and
with a global adjustable phase $\theta$, which allows us to
measure the local fluctuations $\delta \hat{I}(\theta, x)$ of a
given quadrature component. Reference \cite{Treps} gives the
mathematical expressions, in terms of the Green's functions
$G(x,x')$ and $H(x,x')$, of these quantities, from which one
easily derives the best local squeezing $min_{\theta}<(\delta
\hat{I}(\theta, x))^2>$ and the covariance between pixels :
\begin{equation}
C(x,x',\theta)={<\delta \hat{I}(\theta, x)\delta \hat{I}(\theta,
x')>} \label{coveq}
\end{equation}

\section{Results : scalar solitons}

Equation (\ref{mean}) describes the propagation of the electric
field envelope of a monochromatic field of frequency $\omega$ in a
planar waveguide made of a Kerr medium, i.e. having an intensity
dependent index of refraction:

\begin{equation}
n=n_0+n_2|U|^2
 \label{Kerr}
\end{equation}
with $k=n_0 \omega /c$ and $\gamma=n_2 \omega /c$.

It can be written in a universal form if one uses a scaling
parameter $\eta$ and the scaled variables $\zeta= \eta z$, $r=x
\sqrt{2 \eta k}$, and its solution is the well known $\zeta$
invariant hyperbolic secant proportional to $1/cosh r$ : the 1D
scalar soliton, which is able to propagate without deformation in
the nonlinear medium.

\subsection{Squeezing on the total beam}
\begin{figure}
    \centerline{\includegraphics[width=8cm]{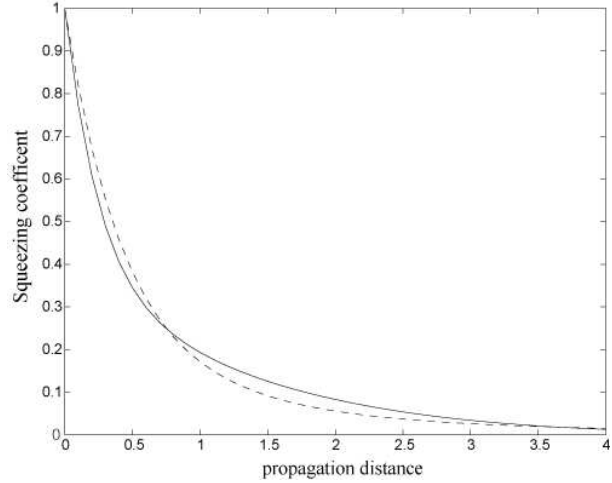}}
    \caption{\label{gsqueez} Best total squeezing, in shot-noise units, observed on the Kerr
     scalar soliton (solid line), versus the normalized propagation distance.
      The dotted line corresponds to the plane wave case. In both cases, the
      normalized propagation distance corresponds to the nonlinear
      phase $\zeta$.}
\end{figure}

As it is well known \cite{Kitagaw}, the Kerr effect produces in
the plane-wave case a significant amount of squeezing, which
increases monotonously with the propagation distance. The
squeezing is optimized for a given quadrature component (best
squeezing) which is neither the amplitude nor the phase
quadrature. However, the amplitude quadrature of the field remains
at the shot noise level. Fig \ref{gsqueez} gives the results of
the Green's function method for the scalar soliton, and shows that
the results are quite similar to the plane wave case, in the case
of a photodetector having an area much larger than the soliton
spot. Neglecting diffraction, or using single mode propagation in
an optical fiber \cite{Levenson, Haus, Friberg, Leuchs} seems
almost equivalent to compensating diffraction by self-focusing,
for a given nonlinear phase $\zeta$. Hence a spatial soliton
appears as a practical mean to obtain strong squeezing, with the
restriction that it must be detected by homodyne techniques.

\subsection{Local quantum  fluctuations.}

\begin{figure}
    \centerline{\includegraphics[width=13cm]{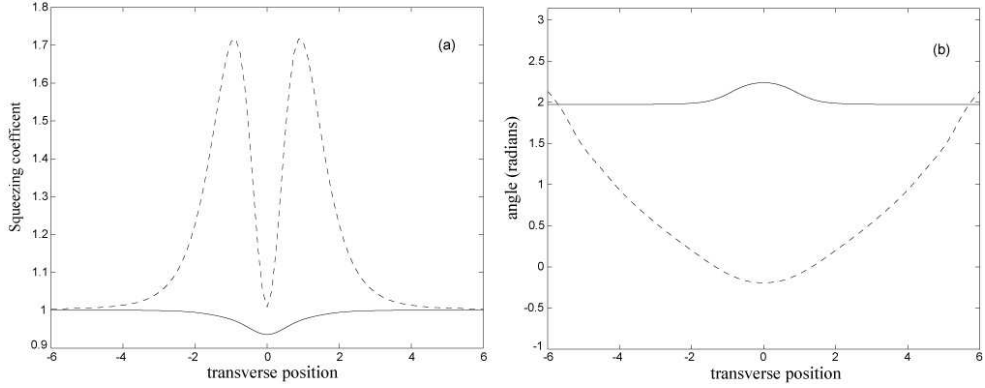}}
    \caption{\label{fig2} (a) Best total squeezing in shot-noise units, and (b) corresponding
    angles, observed on a single pixel, versus
    its transverse position. Solid line : for a propagation distance
    equal to 0.3 Ld. Dashed line : for a propagation distance
    equal to 3 Ld.}

\end{figure}

In order to measure the local quantum fluctuations, we use a
homodyne detector, made of photodetectors of very small area and
quantum efficiency 1, and using a local oscillator at frequency
$\omega$, placed at the output of the nonlinear medium, which is
able to monitor the quantum noise on any quadrature component at
point $(x,z_{out})$ and we vary the local oscillator phase to
reach the minimum noise level : we obtain by this way a quantity
that we call "best squeezing". Figures (\ref{fig2}a) and
(\ref{fig2}b) give, for two normalized propagation distances
($\zeta=0.3$ and $\zeta=3$), such a quantity as a function of x,
together with the phase angle of the local oscillator enabling us
to reach the best squeezed quadrature. One observes that the
central pixel is the most squeezed. However, the noise level
appears to be above the shot noise for the longer propagation
length.

\begin{figure}
    \centerline{\includegraphics[width=8cm]{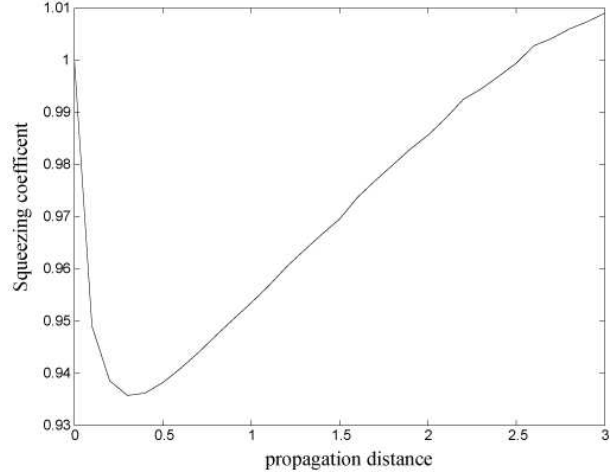}}
    \caption{\label{local}Best total squeezing, in shot-noise units, observed on the central single pixel, versus
    the propagation distance.}

\end{figure}
Indeed, figure \ref{local}, which displays the variation of the
best squeezing on the central pixel as a function of propagation
length, shows that the fluctuations on this pixel decrease until
an optimal propagation length ($\zeta=0.3$), then increase and
overpass the shot noise for a sufficiently long propagation
length. Hence, long propagation lengths produce at the same time a
local excess noise and a squeezed total beam (see fig.1). These
results are not contradictory because, as we will soon see, there
is a gradual build-up of anti-correlations between the different
transverse parts of the soliton  because of diffraction.
\begin{figure}
\centerline{\includegraphics[width=13cm]{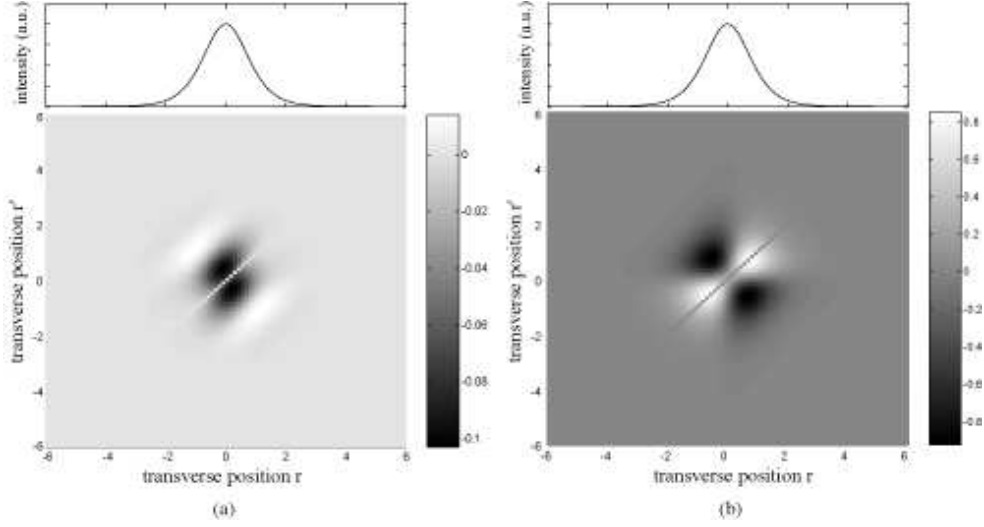}}
    \caption{\label{cov} bottom: Covariance $C(x,x',\theta)$ between
    pixels on the best squeezed quadrature for (a) a
    propagation distance of 0.3 Ld and (b) a propagation distance of 3 Ld. The values on the main diagonal
(variance) have been removed. Top : Corresponding transverse
intensity profile of the scalar soliton.}
\end{figure}
 Figure \ref{cov} shows that the covariance between pixels on the best
squeezed quadrature (calculated with eq.\ref{coveq} and where best
squeezing is defined for the total beam) increases and spreads out
when passing from $\zeta=0.3$ to $\zeta=3$. Indeed, only pixels
close to each other are anti-correlated for $\zeta=0.3$ while the
correlation between adjacent pixels becomes positive for
$\zeta=3$. For this latter distance, squeezing is due to the
anti-correlation between the left and the right side of the
soliton.

\begin{figure}
    \centerline{\includegraphics[width=8cm]{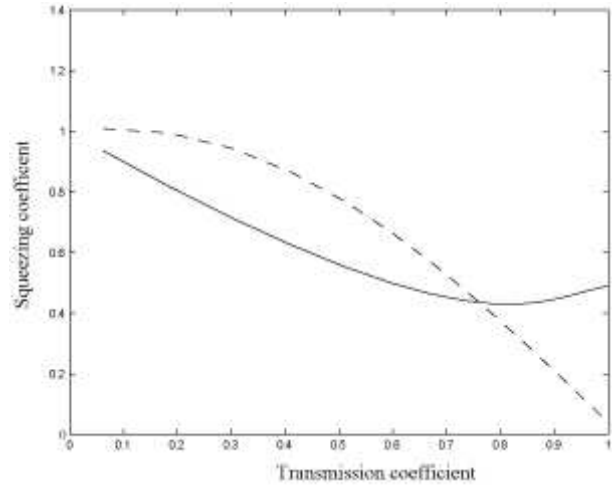}}
    \caption{\label{varsiz}Best squeezing for a photodetector of variable
size, versus the transmission coefficient defined as the ratio
between the detected and the total intensity. Solid line : for a
propagation distance of 0.3 Ld. Dashed line : for a propagation
distance of 3 Ld.}
\end{figure}

Figure \ref{varsiz} shows the best squeezing for a photodetector
of variable size, centered on the beam. This photodetector can be
seen as an iris that allows the detection of only the central part
of the beam, with a minimum size corresponding to the central
pixel and a maximum size corresponding to the detection of the
entire beam. As expected, the noise on a very small area is close
to the shot noise. The squeezing on a small central area is bigger
at small distance, while the squeezing for the entire beam is
better at great distance. For $\zeta=0.3$ the squeezing is maximum
for a finite size of the photodetector (transmission coefficient
of 0.8, corresponding to a normalized radius of 1), while
detecting the entire beam ensures the maximum squeezing for
$\zeta=3$.

In many circumstances, a spatial soliton behaves as a single mode
object. The present analysis allows us to check such a single mode
character at the quantum level : let us assume that the system is
in a single mode quantum state. This means that it is described by
the state vector $|\Psi>\otimes|0>\otimes ...\otimes |0> ...$,
$|\Psi>$ being a quantum state of the mode having the exact
spatial variation of the soliton field, and all the other modes
being in the vacuum state. It has been shown in \cite{martinelli}
that if a light beam is described by such a vector, then the noise
recorded on a large photodetector with an iris of variable size in
front of it has a linear variation with the iris transmission. We
see in figure \ref{varsiz} that it is not strictly speaking the
case : as far as its quantum noise distribution is concerned, a
spatial soliton is not a single mode object. Let us also notice
that the departure from the linear variation is on the order of
$10\%$, so that the single mode approach of the soliton remains a
good approximation.

\subsection{Intensity squeezing by spatial  filtering}
\begin{figure}
    \centerline{\includegraphics[width=8 cm]{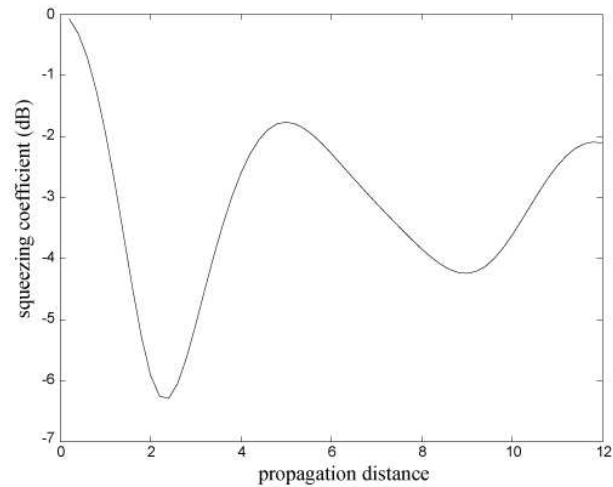}}
    \caption{\label{fft} : intensity squeezing, in dB, for a photodetector
    placed at the center of the Fourier plane, with a total width of
    0.25 in units of spatial frequencies corresponding to the
    normalized units of the direct space.}

\end{figure}
In all  results presented in the preceding  paragraph, squeezing
can be measured only by homodyne detection, because no squeezing
is present for the amplitude quadrature, which is the easiest to
measure, as it requires only direct detection, and no
interferometer. Actually, Mecozzi and Kumar \cite{Mecozzi} have
shown that a simple stop on the central part of the soliton beam
ensures intensity squeezing on the remaining of the light. They
have proposed the following intuitive explanation :  if the
intensity fluctuation is positive for the entire beam, the soliton
becomes narrower (its width is inversely proportional to its
power) and the stop will produce larger losses. In the Fourier
plane, a square diaphragm (low-pass filter) will produce an even
better squeezing. Figure \ref{fft} shows the intensity squeezing
we have obtained with the Green's function method by considering
an aperture placed in the Fourier plane. These results are in full
agreement with the fig. 1 of Ref. 3, that uses another
linearisation method. The residual differences come probably from
a small difference in the aperture width. To conclude this
paragraph, we find, as Mecozzi and Kumar, that subpoissonian light
can be very easily produced from a spatial soliton by placing a
simple aperture in the Fourier plane.

\section{Quantum limits of stability and squeezing on vector solitons}

\subsection{Multimode vector solitons and symmetry-breaking instability}

\begin{figure}
    \centerline{\includegraphics[width=13cm]{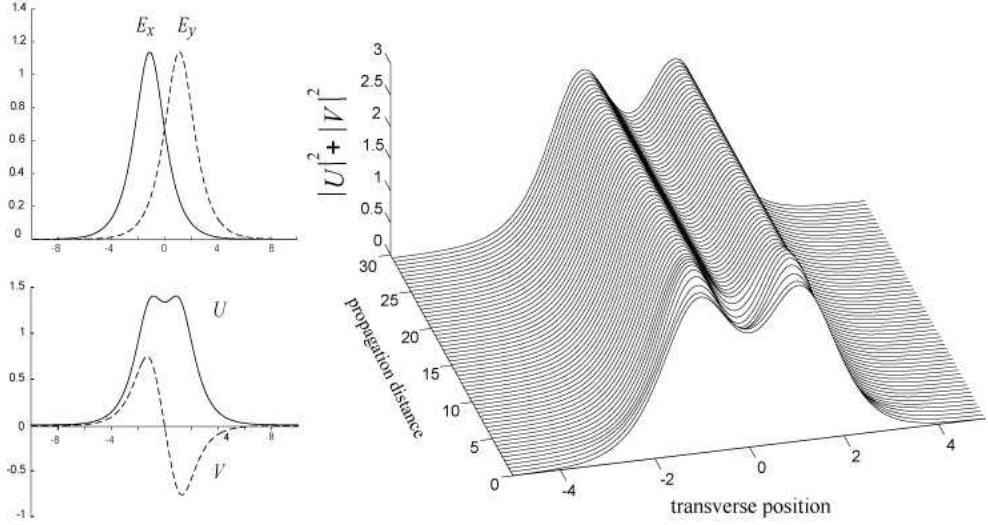}}
    \caption{\label{vect} Vector soliton projected either on orthogonal linear
    polarisations ($E_x$ and $E_y$) or on counter-rotating circular
    polarisations (U and V). The right plot shows invariant propagation
     of the soliton bound-state.}
\end{figure}

In the last few years, a great number of multi-component solitons
have been demonstrated~\cite{kivshar}. Among these, the simplest
example is the bimodal vector soliton which is depicted in fig.
\ref{vect}a and consists of two opposite circular components that
trap each other in a non birefringent Kerr medium. Their complex
envelopes U and V obey the following coupled nonlinear
Schr\"{o}dinger equations :

\begin{equation}
\cases{\frac{\partial U}{\partial Z}=i\frac{1}{2k}
\frac{\partial^2
U}{\partial x^2}+i\gamma(|U|^2U+7|V|^2U)\\
\frac{\partial V}{\partial Z}=i\frac{1}{2k} \frac{\partial^2
V}{\partial x^2}+i\gamma(|V|^2V+7|U|^2V)\\} \label{cnlse}
\end{equation}

\begin{figure}
    \centerline{\includegraphics[width=13cm]{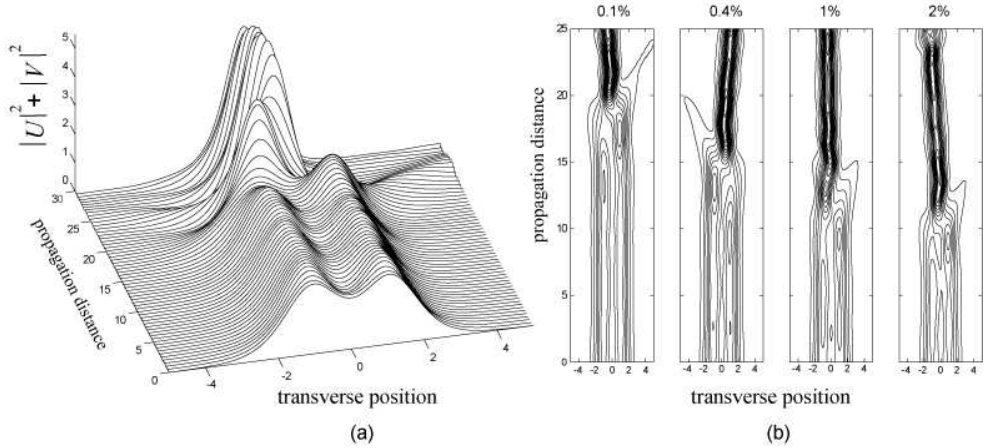}}
    \caption{\label{break} (a) Symmetry-breaking instability of the bimodal soliton represented in
    3-D., and (b) the same in level curves for different initial asymmetry levels
    at the input}

\end{figure}

Where $\gamma$ is the usual Kerr coefficient, and 7 represents the
strength of the cross-phase modulation in an isotropic liquid,
like $CS_{2}$ \cite{Boyd}. As can be seen on fig.\ref{vect}b, this
soliton bound-state propagates in equilibrium due to incoherent
coupling between both circular components.  If alone in the Kerr
medium, the symmetrical field U tends to self-focus. However, the
antisymmetrical field V on the counter-rotating polarization tends
to diffract and it can be shown that a proper choice of the
intensities (in dimensionless units on Figs \ref{vect}) ensures an
exact balance between diffraction and self-focusing. However, this
equilibrium is unstable if the cross-phase modulation coefficient
is greater than the self-phase modulation coefficient. Figure
\ref{break}a displays the solution of the classical equations of
propagation when a very small spatial initial asymmetry is
introduced in one of the two fields. It shows that the multimode
soliton is unstable and eventually collapses, the intensity evenly
distributed between the two cores of the induced waveguide going
abruptly from one core to the other. The instability occurs after
a propagation distance that decreases as the initial asymmetry
increases, as it can be clearly seen in fig. \ref{break}b. A more
complete description of this phenomenon is given in
Ref.~\cite{Kockaert}, while the first experiment demonstrating
symmetry-breaking instability is reported in
Ref.~\cite{Cambournac}. In this experiment, performed with pulses
issued from Nd-Yag at a 10 Hz repetition rate, symmetry-breaking
induced by the noise at the input occurred for about $60 \%$ of
the laser shots, with a left-right repartition equally and
randomly distributed.

\subsection{Stability of vector solitons : quantum limits}

Propagation of a multimode vector soliton in a medium with strong
cross phase-modulation is unstable, as shown theoretically by
Kockaert et al. \cite{Kockaert}, using a linear stability
analysis. Because of quantum noise, symmetry-breaking will occur
even for an ideal symmetrical input beam. The simplest way to
quantitatively assess the phenomenon is propagating a vector
soliton, using Eq.(\ref{cnlse}) and the usual split-step Fourier
method, after the addition on the input fields of a Gaussian white
noise with a variance of half a photon per pixel on each
polarization. The validity of such type of stochastic simulation
is well known in the framework of the Wigner formalism
\cite{Brambilla}. Results show symmetry-breaking with a random
direction after 60 to 70 diffraction lengths.

\begin{figure}
    \centerline{\includegraphics[width=8cm]{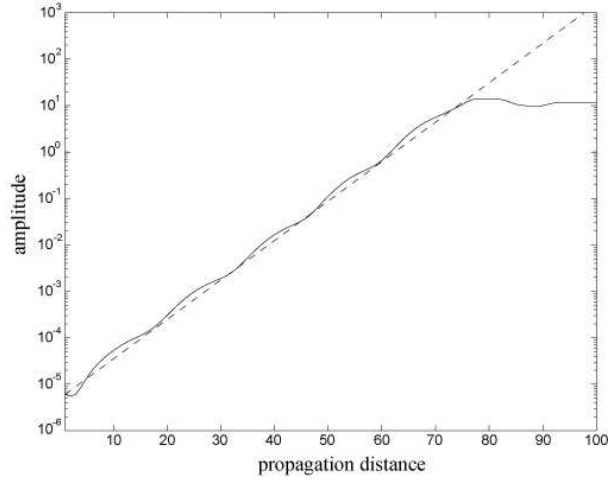}}
    \caption{\label{lstab} Solid line : norm of the difference between the
    unperturbed and the perturbed field, versus the propagation
    distance. The input perturbation corresponds to the unstable
    eigenvector. Dashed line : exponential amplification with the
    corresponding eigenvalue.}
\end{figure}

A linear stability analysis \cite{Kockaert} shows that exponential
amplification occurs only for the part of the input noise obtained
by projection on the eigenvectors associated to eigenvalues with a
negative imaginary part. Such an eigenvector corresponds
physically to a transverse distribution of field for each circular
polarization. The part of an unstable eigenvector that is
circularly polarized as the U symmetrical field is
antisymmetrical, while the part of the eigenvector that is
circularly polarized as the V antisymmetrical field is
symmetrical. Moreover, exponential amplification corresponding to
the eigenvalue having the most negative imaginary part is
predominant after a short propagation distance. Hence, instability
can be described as the amplification of a single unstable
eigenvector. In the results of figure \ref{lstab}, the field
corresponding to this eigenvector with an energy of half a photon
per polarization has been added at the input to the ideal vector
soliton.  Figure \ref{lstab} shows in solid line the evolution of
the norm of the difference between the input field and the
propagated field, compared to a pure exponential amplification
with the corresponding imaginary eigenvalue (dashed line). The
difference evolves almost exponentially, with however a periodic
modulation due to the real part of the eigenvalue. This phase
modulation leads to a dependence between the direction of
symmetry-breaking and the initial amplitude of the perturbation.
Symmetry-breaking appears clearly on figure \ref{lstab} when the
initial small perturbation has grown until a macroscopic level,
for a propagation distance of about 70 diffraction lengths.

To conclude this paragraph, the usual classical simulation of
quantum noise allows us to predict a maximum propagation length
for the multimode vector soliton. This length is less than two
meters, for the diffraction length characterizing the experiment
of ref.\cite{Cambournac}, but also 70 times the typical length
leading to symmetry-breaking in this experiment. Actually,
experimental evidence of quantum limits for symmetry-breaking
appears as a rather difficult challenge, because unavoidable
spatial classical noise dominates in realistic input fields.

\subsection{\label{vbeam}Total beam squeezing and correlation between
polarisations}

In this paragraph, we give the results of the Green's function
method in the case of vector solitons. Because these solitons are
not analytically integrable, the impulse response due to a
Dirac-like field modification of the perfect vector soliton is
more complicated to determine, and calculated as follows. The
perturbation consists in a unity single-pixel step, multiplied by
a small coefficient before to be added to the field in order to
ensure a near-perfect linearity of propagation equations for the
perturbation. Both the unmodified and the modified field are first
numerically propagated using eq.\ref{cnlse}, then subtracted each
other. Finally, the subtraction result is divided by the initial
multiplication coefficient, in order to retrieve the output
corresponding to the input unity single-pixel step. An important
point is that the phase reference corresponding to the amplitude
quadrature is given for each pixel by the unmodified field.
Indeed, in practice the vector soliton could experience a very
slight breathing, that results in a weak phase curvature.
Therefore, taking the mean soliton phase as reference leads to
strong numerical inaccuracies for strong squeezing.
\begin{figure}[!h]

    \centerline{\includegraphics[width=13cm]{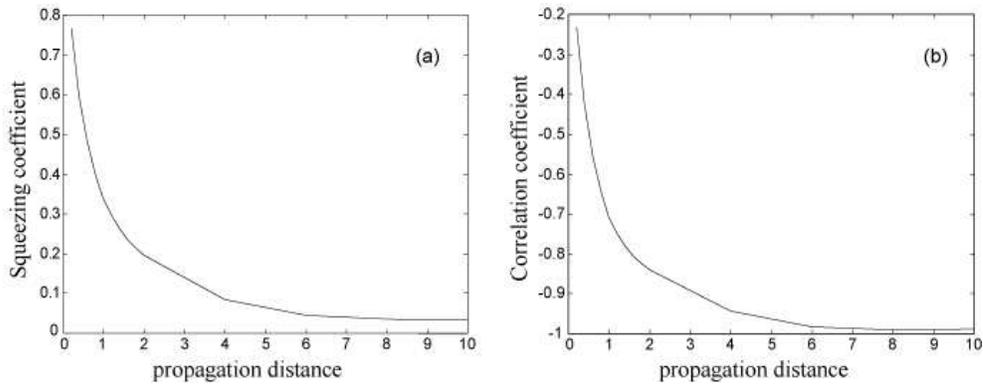}}
    \caption{\label{glob} : best total squeezing in shot-noise units
    (a) and correlation coefficient between circular
    polarisations
    (b) versus the normalized propagation distance.}

\end{figure}

\begin{figure}
    \centerline{\includegraphics[width=8 cm]{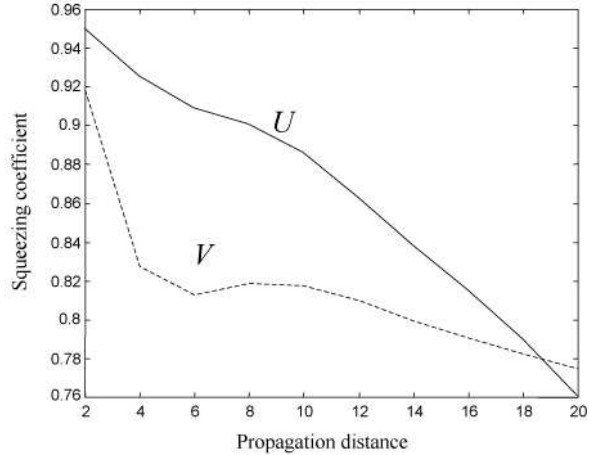}}
    \caption{\label{gcirc} : best squeezing in shot-noise units
    on each circular polarisation
     versus the normalized propagation distance.}

\end{figure}

Fig \ref{glob} shows the global squeezing coefficient of the
vector soliton on the best quadrature. This figure is very similar
to that obtained for a scalar soliton (figure \ref{gsqueez}) :
strong squeezing occurs, although somewhat smaller than in the
scalar case for the same propagation distance. However, this
squeezing is due to a strong anti-correlation between circular
components, as shown in figure \ref{glob} b. It can be verified on
figure \ref{gcirc}, which displays the best squeezing of each
circular polarisation component as a function of propagation, that
the best squeezing on a single circular polarization is much
weaker. We can conclude that a vector soliton as a whole exhibits
quantum properties that are similar to a scalar soliton, while the
field on each polarisation does not constitute itself a soliton.

\subsection{Local quantum fluctuations}
\begin{figure}[!h]
    \centerline{\includegraphics[width=13 cm]{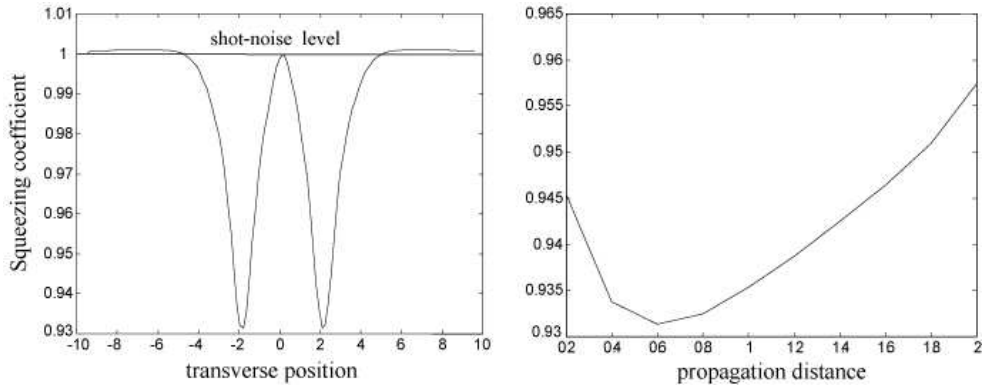}}
    \caption{\label{vlocal}  best squeezing on one pixel.
   a : versus the transverse position for $\zeta$ = 0.6. b : versus the
   propagation distance for the transverse position corresponding
   to the strongest squeezing}

\end{figure}

\begin{figure}[!h]
    \centerline{\includegraphics[width=8 cm]{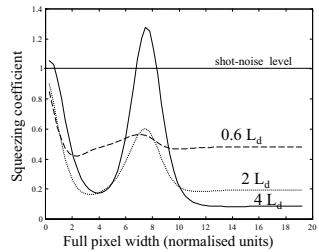}}
    \caption{\label{vsiz}  Best squeezing for a photodetector centered on one intensity peak
    and of variable size, for 3 different propagation distances, versus the transmission
coefficient defined as the ratio between the detected and the
total intensity.}

\end{figure}

Figure \ref{vlocal} shows the squeezing coefficient on one pixel
versus its transverse position, for the propagation length where
this local squeezing is the most intense. The degree of squeezing
is similar to the scalar case (see figure \ref{fig2}), with
however two peaks of squeezing located on the intensity peaks of
the multimode vector soliton. As in the scalar case, local
squeezing is maximum for a relatively small propagation length
(fig.\ref{vlocal} b). Figure \ref{vsiz} shows the best squeezing
for a photodetector of variable size, centered on one of the
intensity peaks. In contrast to figure (\ref{varsiz}) the
variation is far from being linear with the transmission, meaning
that the system is not single mode, as expected. One observes that
for long propagation distances, the global squeezing reaches a
maximum for a photodetector size corresponding roughly to the peak
width. This global squeezing implies the existence of spatial
anti-correlations. Indeed, the local squeezing on one pixel
disappears for such long distances (see figures \ref{vlocal} and
\ref{vsiz}).

\begin{figure}[!h]
    \centerline{\includegraphics[width=13cm]{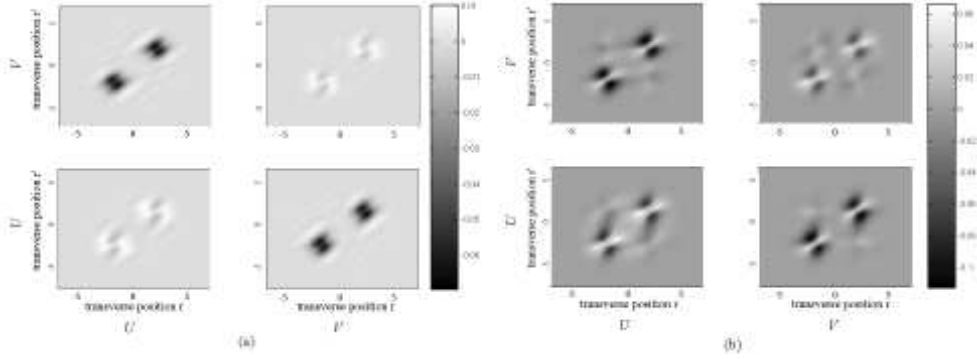}}
    \caption{ \label{vcor}Covariance $C(x,x',\theta)$, in shot-noise units, between
    pixels on the best squeezed quadrature for two propagation
    distances (a) 0.6 Ld and (b) 2 Ld. The values on the main diagonal (variance) have
    been removed.}

\end{figure}

The correlations between fluctuations at different pixels, and
possibly on different polarizations, provide a great deal of
information about the distribution of fluctuations in the soliton.
Figure \ref{vcor} shows these spatial
 correlations, in terms of
the covariance functions $C(x,x',\theta)$ of the two
polarizations, and the covariance function between the two modes,
on the best squeezed quadrature and for 2 propagation lengths. One
sees that for $\zeta=0.6 Ld$, anti-correlations appear between
pixels when one measures different circular polarizations on the
pixels. It means that the effect of cross-phase modulation is
already noticeable, while effects of diffraction and self-phase
modulation are still weak. For $\zeta=2 Ld$, anticorrelations
appear also between pixels on the same polarization, as for the
scalar soliton.

\subsection{Fluctuations on linear polarisations}

\begin{figure}[!h]
    \centerline{\includegraphics[width=13 cm]{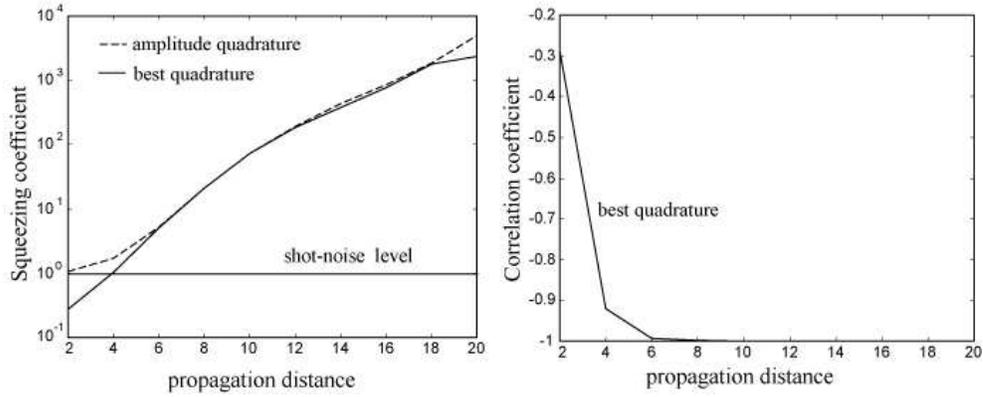}}
    \caption{\label{vlin}  Squeezing coefficient (a) and correlation coefficient (b) of linear
    polarisations
    versus the  propagation
    distance.}
\end{figure}

Figure \ref{vlin} shows that squeezing on linear polarisations
disappears after a relatively short propagation distance, even on
the best quadrature. Moreover, fluctuations grow exponentially. On
the other hand,  fluctuations on both linear polarisations are
perfectly anti-correlated after some distance, as it was expected
because of the strong squeezing of the global beam. We have
verified that the results on this global beam (section
\ref{vbeam}) can be retrieved also from a description of the field
by projection on linear polarisations.

\section{Conclusion}

Let us summarise here the main results about the spatial
distribution of quantum fluctuations in 1D Kerr solitons that we
have obtained in this paper :  as far as measurements on the whole
soliton are considered, we have shown that scalar Kerr solitons,
as well as vector Kerr solitons, exhibit squeezing properties
similar to that of a plane wave propagating in the same medium and
for the same type of propagation distance. Because of diffraction,
as the soliton propagates, strong anti-correlations develop
between symmetrical points inside the scalar soliton spot, and
between  polarisation components for the vector soliton.

\section{Acknowledgments}

This work has been supported by the European Union in the frame
of the QUANTIM network (contract IST 2000-26019).


\section{References}



\begin{thebibliography}{12}


\bibitem{kolobov} M.I. Kolobov, Rev.  Mod.  Phys.  \textbf{71},1539
(1999).

\bibitem{Nagasako} E.M. Nagasako, R.W. Boyd, G.S. Agrawal,
Optics Express \textbf{3} 171-179 (1998)


\bibitem{Mecozzi} A. Mecozzi, P. Kumar, Quantum and Semiclassical
 Optics \textbf{10} L21-L26 (1998)


\bibitem{Treps} N. Treps and C. Fabre, Phys. Rev. A \textbf{62}
033816 (2000)

\bibitem{Lantz} E. Lantz, N. Treps, C. Fabre, E. Brambilla,
"Spatial distribution of quantum fluctuations in spontaneous
down-conversion in realistic situations : comparison between the
stochastic approach and the Green's function method." , submitted

\bibitem{Kockaert} P. Kockaert, M. Haeltermann, J. Opt. Soc. Am. B
\textbf{16} 732-740 (1999)

\bibitem{Cambournac} C. Cambournac, T. Sylvestre, H. Maillotte, B. Vanderlinden,
 P. Kockaert, Ph. Emplit, and M. Haelterman, Physical Review Letters \textbf{89} 083901 (2002)

\bibitem{Levenson} R. Shelby, M. Levenson, S. Perlmutter, R. De
Voe, D. Walls, Phys. Rev. Lett. \textbf{57} 681 (1986)

\bibitem{Haus}K. Bergman, H. Haus, Optics Lett. \textbf{16} 663
(1991)

\bibitem{Friberg} S. Friberg, S. Machida, Y. Yamamoto, Phys. Rev. Lett. \textbf{69} 3775 (1992)

\bibitem{Leuchs} S. Spalter, N. Korolkova, F. Konig, A. Sizmann, and G.
Leuchs, Phys. Rev. Lett. \textbf{81} 786 (1998)

\bibitem{Kitagaw} M. Kitagawa, Y. Yamamoto, Phys. Rev. A \textbf{34} 3974 (1986)

\bibitem{martinelli} M. Martinelli, N. Treps, S. Ducci, S. Gigan, A. Maitre, C. Fabre,
 Phys. Rev. A \textbf{67} 023808 (20003)

\bibitem{kivshar} Y.S. Kivshar and G.P. Agrawal, \emph{Optical solitons
: From fibers to Photonic Crystals}, (Academic Press, San Diego,
 2003).


\bibitem{Boyd} R.W. Boyd, \emph{Nonlinear optics}, (Academic Press, San Diego,
1992).

\bibitem{Brambilla}E. Brambilla, A.Gatti, M. Bache and  L.A.Lugiato, "Simultaneous
near-field and far field spatial quantum correlations in
spontaneous parametric down-conversion", submitted.

\end{thebibliography}
\end{document}